\documentclass[preprintnumbers,amsmath,amssymb,superscriptaddress, showpacs,twocolumn]{revtex4-1}%

\usepackage{hyperref}
\hypersetup{
  colorlinks=true,        
  linkcolor=blue,         
  citecolor=cyan,         
}

\usepackage{enumitem}
\usepackage{graphicx,subfigure}
\usepackage{dcolumn}
\usepackage{bm}
\usepackage{color}

\newcommand{\be}{\begin{eqnarray}}
\newcommand{\ee}{\end{eqnarray}}

\begin{document}

\title{Weak gravitational lensing: a compact object with arbitrary quadrupole moment immersed in plasma}

\author{Hrishikesh~Chakrabarty}
\affiliation{Center for Field Theory and Particle Physics and Department of Physics, Fudan University, 200438 Shanghai, China}

\author{Askar B. Abdikamalov}
\affiliation{Center for Field Theory and Particle Physics and Department of Physics, Fudan University, 200438 Shanghai, China}

\author{Ahmadjon A. Abdujabbarov}
\affiliation{Center for Field Theory and Particle Physics and Department of Physics, Fudan University, 200438 Shanghai, China}

\affiliation{Ulugh Beg Astronomical Institute,
Astronomicheskaya 33, Tashkent  100052, Uzbekistan}

\author{Cosimo~Bambi}
\email[Corresponding author: ]{bambi@fudan.edu.cn}
\affiliation{Center for Field Theory and Particle Physics and Department of Physics, Fudan University, 200438 Shanghai, China}
\affiliation{Theoretical Astrophysics, Eberhard-Karls Universit\"at T\"ubingen, 72076 T\"ubingen, Germany}

\date{\today}

\begin{abstract}
We study weak gravitational lensing around a compact object with arbitrary quadrupole moment in the presence of plasma. The studied compact objects are considered to be spherically symmetric. The additional parameter $\epsilon$ regulating the quadrupole moment in the metric alters the deflection angle of light rays along with the plasma parameters. In the vacuum, the number of images due to the presence of the parameter $\epsilon$ increases and causes the increase of the magnification of the image source. The effects of uniform and nonuniform plasma on gravitational lensing around the compact object are also studied. 
\end{abstract}

\maketitle


\section{Introduction}

Gravitational lensing is one of the distinguishable features of the general relativistic theory of gravity. Light propagating near a compact object will be deflected due to gravitational effects (for the review, see e.g.~\cite{Synge60, Schneider92, Perlick00}).  On the other hand, almost all gravitational lensing objects are surrounded by plasma and the interstellar medium. For astrophysical applications, it is thus very important and interesting to study the impact of plasma on gravitational lensing. The effect of uniform and nonuniform plasma on light propagation near compact gravitational objects have been considered by various authors~\cite{Rogers15,Rogers17,Er18,Rogers17a,Broderick03, Bicak75, Kichenassamy85, Perlick17, Perlick15, Abdujabbarov17, Abdujabbarov17b, Eiroa12, Kogan17}. Another optical property related to light propagation is the shadow of black holes and other compact objects, and it has also been considered in the literature~\cite{Abdujabbarov17c, Abdujabbarov16a, Abdujabbarov16b, Abdujabbarov15, Abdujabbarov15b, Abdujabbarov15a, Atamurotov15a, Amarilla12, Bambi09, Bambi15, Ghasemi-Nodehi15, Cunha15, Wei13, Takahashi05, Ohgami15, Nedkova13, Falcke00, Grenzebach15, Takahashi04, Li14a, Wei15b, Papnoi14, Bambi10, Atamurotov13b, Atamurotov13, Tsukamoto14, Grenzebach2014, Abdujabbarov13c, Amarilla10, Amarilla13, Hioki09, Li2014}.

Microlensing, or weak gravitational lensing, is one of the most attractive types of lensing. When one considers microlensing, it is very important to note that the images of the sources are optically unresolved. However, the effect can be observed due to the magnification of the brightness of the radiating object.

First studies on microlensing have been reported, for instance, in~\cite{Paczynski86a, Alcock93, Aubourg93, Udalski93, Paczynski96}. For a review on gravitational lensing in general relativity, see~\cite{Perlick04}. At the same time, strong gravitational lensing around spherically symmetric compact objects is described in~\cite{Tsupko09}. The work~\cite{Tsupko09} has been dedicated to studying the angular sizes and magnification factors for relativistic rings formed by the photons undergoing one or several turns around a black hole. The plasma effects on gravitational lensing have been studied, e.g., in~\cite{Kogan10,Tsupko10,Tsupko12,Morozova13,Tsupko14,Shaikh17,Shaikh18a,Shaikh18b}.

Astrophysical black holes are described mainly by their mass and rotation parameter expressed by the Kerr solution. However, axial symmetric black holes, in principle, can have more parameters. For example, many authors have studied the possibility of testing rotating black holes with non-vanishing electric charge~\cite{Grunau11,Zakharov94,Stuchlik02, Pugliese10, Pugliese11b, Pugliese11, Patil12}, black holes with brane charge~\cite{Turimov17, Whisker05, Majumdar05, Liang17, Li15}, black holes with gravitomagnetic charge~\cite{Liu11,Zimmerman89, Morozova09, Aliev08, Ahmedov12, Abdujabbarov11, Abdujabbarov08}, and deformation parameters of parametrized axial symmetric metrics~\cite{Bambi17c,Rayimbaev15, Bambi11b, Chen12, Bambi12}. One of the possible deviations from the Kerr solution has been proposed by Glampedakis and Babak~\cite{Glampedakis06b} using an approximate solution of Einstein vacuum equations and adding the leading order deviation which appears in the value of the spacetime quadrupole moment. Different physical properties of these quasi-Kerr black holes have been studied in~\cite{Psaltis12, Liu12b}.

In this paper, we consider the weak gravitational lensing near a spherically symmetric compact objects with arbitrary quadrupole moment. We also consider the presence of plasma around the lensing object and study both the deformation parameter and the plasma characteristics on photon motion, lensing effect, and image source magnification.

The paper is organised as follows. %
The light propagation in this spacetime in the presence of plasma and the weak lensing effect around quasi-Kerr black holes are considered in Sect.~\ref{weaknr}. The magnification of the image source due to weak lensing is considered in Sect.~\ref{sect:magnification}.
We conclude our results in Sect.~\ref{sec:conclusion}. In Appendix~\ref{s-review}, we briefly review the quasi-Kerr metric. 
In the present paper, we adopt a metric with signature $(-,+,+,+)$ and we employ a system of geometric units in which $G = 1 = c$. Greek indices run from 1 to 3. Latin indices run from 0 to 3.


\section{Weak lensing in the presence of plasma}\label{weaknr}

For our calculations, we consider an approximation of the metric~(\ref{qkmetric}) which is described in the Appendix~\ref{s-review}.

\subsection{Expression for deflection angle}
In this subsection, we investigate the effect of a plasma on the deflection of light around a compact object, which is an approximated solution in the weak field limit. For that purpose, we consider the non-rotating case ($ a=0 $). For the calculation of the deflection angle in an inhomogeneous plasma in the presence of gravity, we follow the derivation in~\cite{Kogan10, Tsupko10,Tsupko12,Morozova13,Tsupko14, Synge60}. 

A non-vanishing quadrupole moment will give rise to some deviations from spherical symmetry but we employ the conditions $ \theta=\pi/2 $,  $ a=0 $, and the weak-field approximation. Under these approximations, we consider the metric to be spherically symmetric. We write the line element~(\ref{qkmetric}) as 
\begin{equation}\label{wfapprx}
g_{ik}=\eta_{ik}+h_{ik}, \ \ \ \ | h_{ik} | \ll1.
\end{equation}
Here, $ \eta_{ik} $ is the flat spacetime metric and $ h_{ik} $s' are small perturbations. For the contravariant components, we have~\cite{Landau-Lifshitz2, Kogan10}
\begin{equation}
g^{ik}=\eta^{ik}-h^{ik}, \ \ \ \ \eta_{ik}=\eta^{ik}, \ \ \ \ h_{ik}=h^{ik}.
\end{equation}

We consider a static and inhomogeneous plasma in this gravitational field whose refractive index is given by
\begin{equation}
n^{2}=1-\frac{\omega_{p}^{2}}{\omega^{2}}, \ \ \ \ \omega_{p}^{2}=\frac{4\pi e^{2}N(x^{\alpha})}{m}=K_{e}N(x^{\alpha}) ,
\end{equation}
where $ \omega $ is the frequency of the photon, which depends on the space coordinates $ x^{\alpha} $, and $ N(x^{\alpha}) $ is the electron density in the inhomogeneous plasma. $ e $ and $ m $ are the electronic charge and mass, respectively, and $ \omega_{p} $ is the plasma frequency. 

The trajectories of light rays in plasma in the presence of a gravitational field can be obtained from the equation~\cite{Synge60}
\begin{equation}{\label{weakham}}
W(x^{i},p_{i})=\frac{1}{2}\left[g^{ij}p_{i}p_{j}+\omega_{p}(x^{\alpha})^{2}\right]=0\ .
\end{equation} 

From (\ref{weakham}) we obtain the following system of equations for the space components $ x^{\alpha} $ and $ p_{\alpha} $
\begin{equation}
\frac{dx^{\alpha}}{d\lambda}=g^{\alpha\beta}p_{\beta}, \ \ \ \ \frac{dp_{\alpha}}{d\lambda}=-\frac{1}{2}g^{ij}_{,\alpha}p_{i}p_{j}-(\omega_{p}^{2})_{,\alpha}\ .
\end{equation}  

Now, using the weak field approximation (\ref{wfapprx}), the equation for the deflection angle is 
\begin{equation}
\hat{\alpha}_{\alpha}=\frac{1}{2}\int_{-\infty}^{\infty}\left(h_{33,\alpha}+\frac{h_{00,\alpha}}{1-(\omega_{p}^{2}/\omega^{2})}-\frac{K_{e}N_{,\alpha}}{\omega^{2}-\omega_{p}^{2}}\right)dz 
\end{equation}
Hereafter, we assume that the photons are moving along the $ z $ direction (which coincides with the axis of symmetry) and the impact parameter $ b $ is introduced for convenience which remains constant in the null approximation for photons moving along the axis $ z $. The plasma has a spherically symmetric distribution around the lens, with concentration $ N=N(r) $. In the axially symmetric situation, the position of a photon can be characterized by $ b $ and $ z $, and the absolute value of the radius vector is $ r=\sqrt{x_{1}^{2}+x_{2}^{2}+z^{2}} = \sqrt{b^{2}+z^{2}} $. Using this simplifications, the expression for the deflection angle becomes
\begin{eqnarray}\label{defangle}
\hat{\alpha}_{b}&=&\frac{1}{2}\int_{-\infty}^{\infty}\frac{b}{r} \bigg(\frac{dh_{33}}{dr}+\frac{1}{1-(\omega_{p}^{2}/\omega^{2})}\frac{dh_{00}}{dr}\nonumber\\ 
&& -\frac{K_{e}}{\omega^{2}-\omega_{p}^{2}}\frac{dN(r)}{dr}\bigg)\ .
\end{eqnarray}

\subsection{Calculation of the deflection angle}

In this subsection, we calculate and study the dependence of the deflection angle on various parameters of the spacetime (\ref{qkmetric}) and of the plasma. For the non-rotating case, in the weak field limit and in Cartesian frame the components $ h_{ik} $s' can be expressed as 
\begin{eqnarray}\label{hij}
h_{00}&=&\frac{2M}{r}-\frac{2 M^{2}}{ r^2}- \frac{\epsilon M^{3}}{r^{3}}, \\ 
h_{\alpha\beta}&=&\left(\frac{2M}{r}-\frac{2 M^{2}}{ r^2}- \frac{\epsilon M^{3}}{r^{3}}\right)s_{\alpha}s_{\beta}\ , \\
h_{33}&=&\left(\frac{2M}{r}-\frac{2 M^{2}}{ r^2}- \frac{\epsilon M^{3}}{r^{3}}\right)\cos^{2}\theta
\end{eqnarray} 
Here $ s_{\alpha} $ is the unit vector in the direction of the radius vector $ r_{\alpha} $, and its components are equal to the direction cosines. The angle $ \theta $ is the angle between $ r_{\alpha}=r^{\alpha} $ and the $ z $ axis. Since $ s_{3}=\cos\theta=z/r=z/\sqrt{b^{2}+z^{2}} $, the component of $ h_{33} $ will have the form
\begin{equation}\label{h33}
h_{33}=\left(\frac{2M}{r}-\frac{2 M^{2}}{ r^2}- \frac{\epsilon M^{3}}{r^{3}}\right)\frac{z^{2}}{b^{2}+z^{2}}\ .
\end{equation}
Let us consider a plasma with an
electron density that decreases as a function of distance, such that~\cite{Rogers15,Rogers17,Abdujabbarov17, Abdujabbarov17b}
\begin{equation}\label{plasmadist}
\omega_{p}^{2}=K_{e}N(r)\ .
\end{equation}
Now, using Eqs.~(\ref{hij}) and (\ref{h33}), we rewrite Eq.~(\ref{defangle}) in the following form
\begin{equation}\label{numint}
\begin{aligned}
&\hat{\alpha}_{b}=\frac{1}{2}\int_{-\infty}^{\infty}\frac{b}{r}\Bigg[ \frac{d}{dr}\left(\left( \frac{2M}{r}-\frac{2 M^{2}}{ r^2}- \frac{\epsilon M^{3}}{r^{3}}\right)\frac{z^{2}}{b^{2}+z^{2}}\right) \\
& \ \ \ \ \ \ +\frac{\omega^{2}}{\omega^{2}-\omega_{p}^{2}} \frac{d}{dr}\left( \frac{2M}{r}-\frac{2 M^{2}}{ r^2}- \frac{\epsilon M^{3}}{r^{3}}\right) \\
& \ \ \ \ \ \ -\frac{K_{e}}{\omega^{2}-\omega_{p}^{2}}\frac{d}{dr}N(r)  \Bigg]dz\ .
\end{aligned}
\end{equation}
Considering the special case when $ \omega_{p}^{2}/\omega^{2}\ll 1 $, we can get information on the physical meaning of each term in Eq.~(\ref{numint}). We obtain,
\begin{equation}\label{numint2}
\begin{aligned}
&\hat{\alpha}_{b}=-\frac{4M}{b}+\frac{3\pi M^2}{b^{2}} +\frac{16\epsilon M^3}{3b^3}\\ 
& \ \ \ \ \ \ \ +\frac{1}{2}\int_{-\infty}^{\infty}\frac{b}{r} \frac{\omega_{p}^{2}}{\omega^{2}} \frac{d}{dr}\left( \frac{2M}{r}-\frac{2 M^{2}}{ r^2}- \frac{\epsilon M^{3}}{r^{3}}\right)dz \\
& \ \ \ \ \ \ \  -\frac{1}{2}\int_{-\infty}^{\infty}\frac{b}{r} \frac{K_{e}}{\omega^{2}}\Big( 1+\frac{\omega_{p}^{2}}{\omega^{2}} \Big) \frac{d}{dr}N(r) dz\ .
\end{aligned}
\end{equation}

\begin{figure*}[t]
	\begin{center}
		\includegraphics[width=0.32\textwidth]{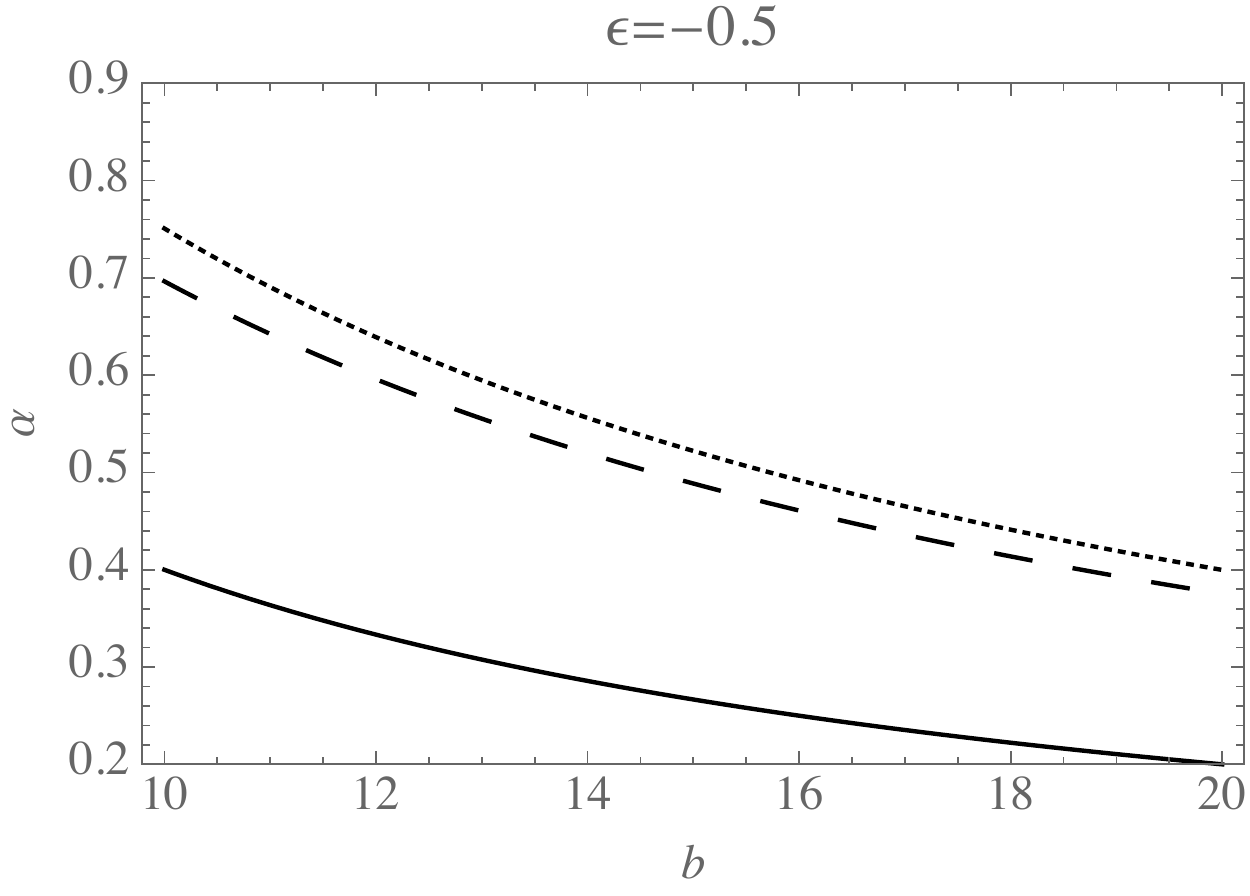}
		\includegraphics[width=0.32\textwidth]{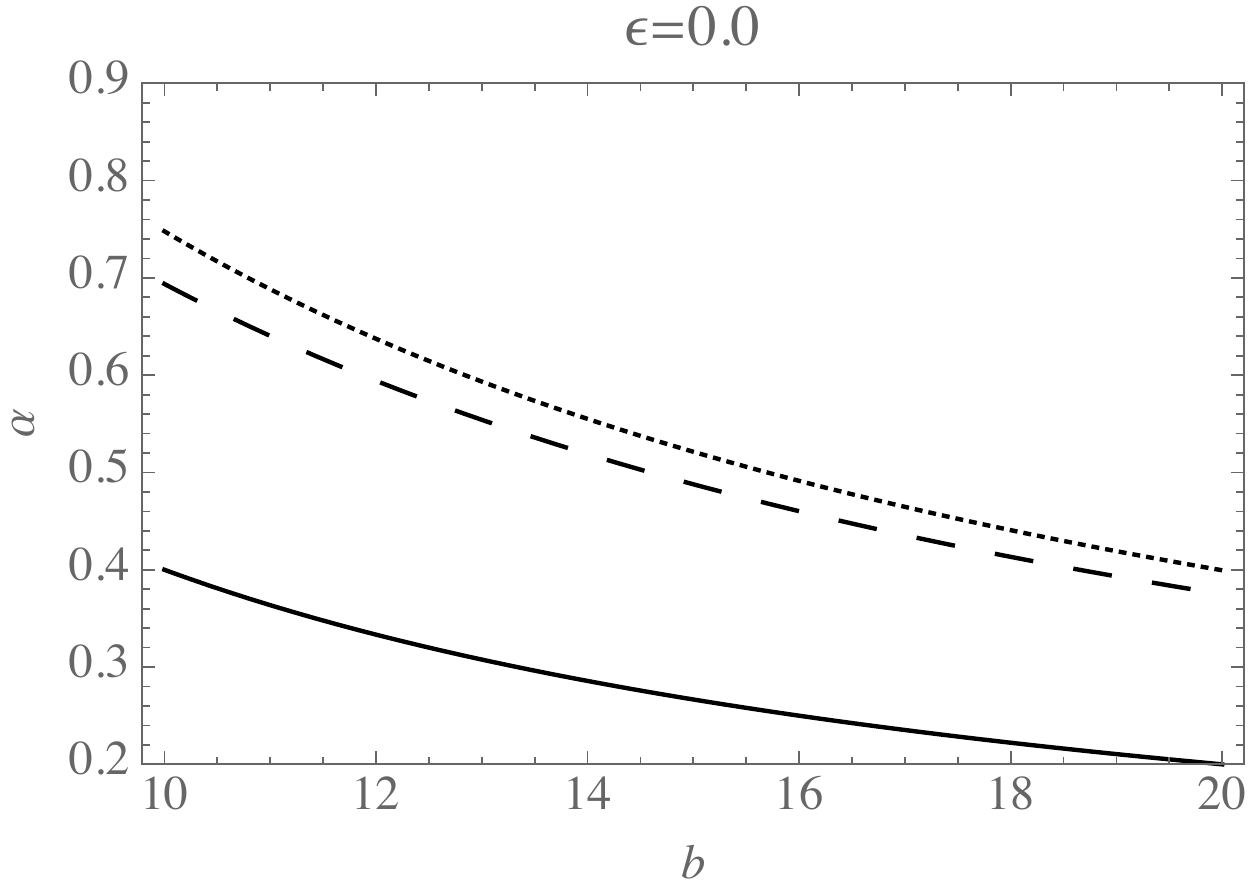}
		\includegraphics[width=0.32\textwidth]{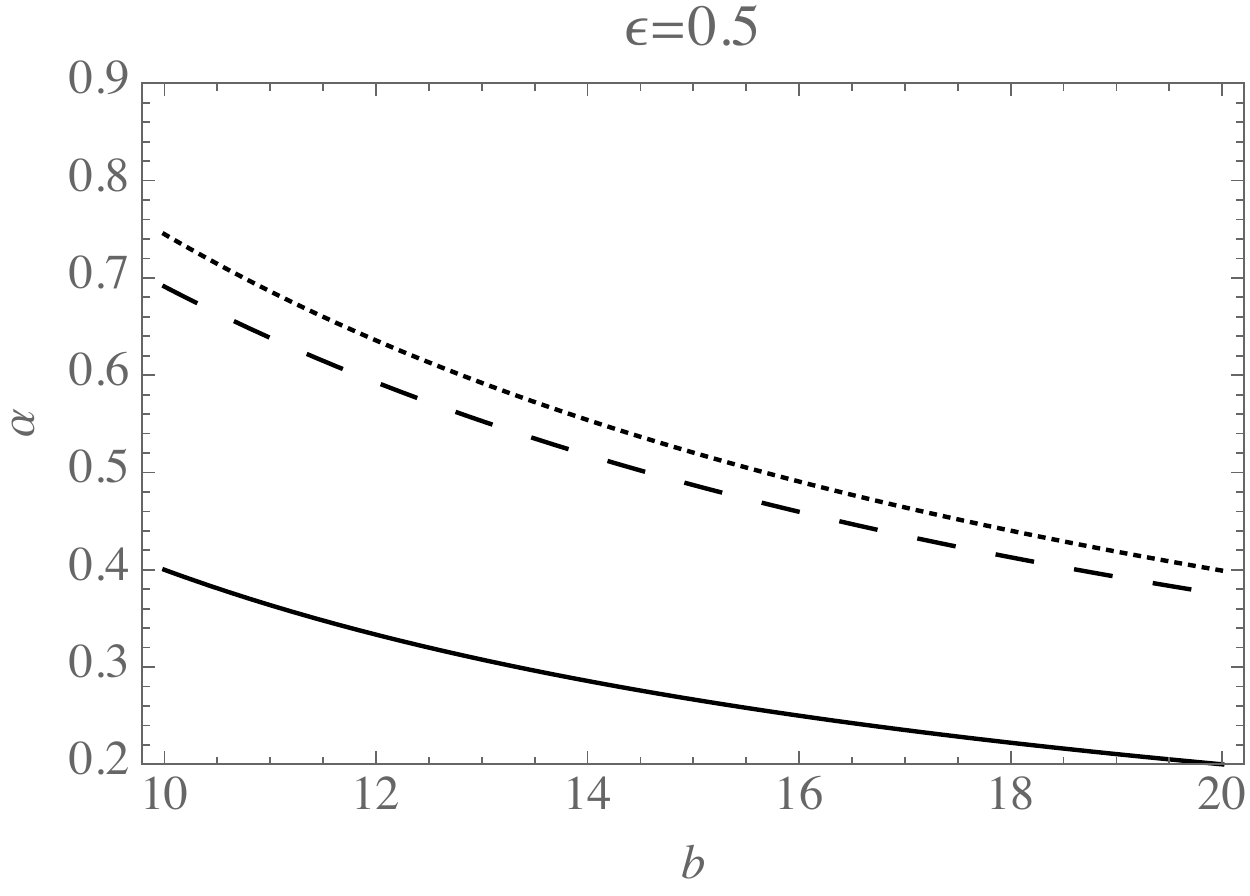}
	\end{center}
	\vspace{-0.5cm}
	\caption{The dashed line corresponds to deflection for non-uniform plasma with power-law electron density $N(r)=N_0/ r^3$. The electron density at the radial coordinate $r=3M$ is set at $ 10^{-4}\ {\rm cm}^{-3} $. The dotted line corresponds to a uniform plasma distribution with the electron density $1.3\cdot 10^{-4}\ {\rm cm}^{-3}$. In all cases, the photon frequency is $4\cdot 10^{10}\ {\rm Hz}$. The solid line corresponds to deflection without plasma and quadrupole correction (Schwarzschild limit).
	\label{defang} }
\end{figure*}

\begin{figure}[t]
	\begin{center}
		\includegraphics[width=0.45\textwidth]{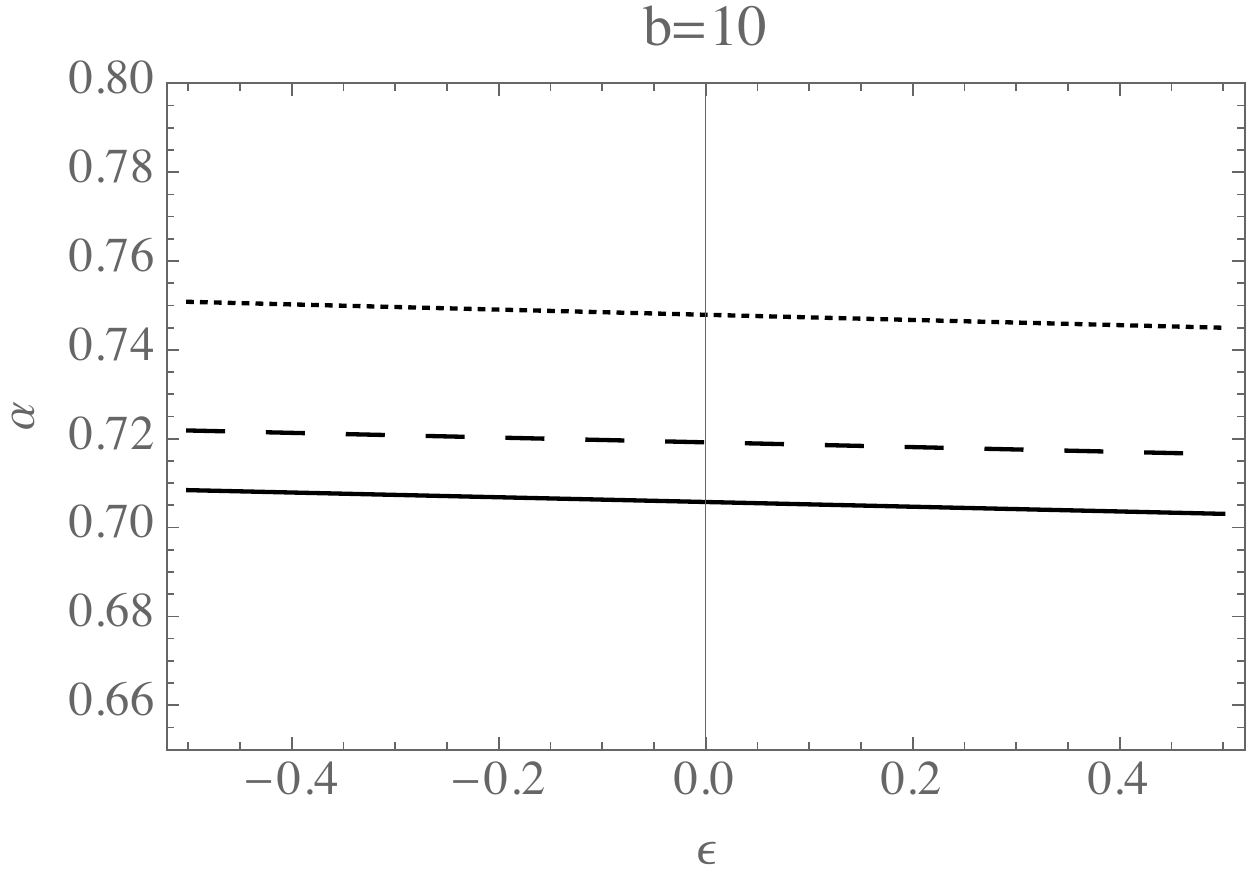}
	\end{center}
	\caption{The dashed line corresponds to deflection for non-uniform plasma with power-law electron density $N(r)=N_0/ r^3$. The electron density at the radial coordinate $r=3M$ is set at $ 10^{-4}\ {\rm cm}^{-3} $. The dotted line corresponds to a uniform plasma distribution with the electron density $1.3\cdot 10^{-4}\ {\rm cm}^{-3}$. The photon frequency is always $4\cdot 10^{10}\ {\rm Hz}$.  The solid line corresponds to deflection without plasma. \label{defang2}}
\end{figure}

In Eq.~(\ref{numint2}), the first term corresponds to the vacuum gravitational deflection. The second and third term corresponds to the correction in the gravitational deflection due to a possible non-zero quadrupole moment. The third term in Eq.~(\ref{numint2}) corresponds to an additive correction to the gravitational deflection due to the presence of plasma; note that this term is present in the expression for the deflection angle in the case of both homogeneous and inhomogeneous plasma and depends on the photon frequency as well (due to dispersion). The fourth term on the right-hand side of Eq.~(\ref{numint2}) is a correction to the third term and is a pure effect of the inhomogeneity of the plasma environment. Fig.~\ref{defang} shows the dependence of the deflection angle on the impact parameter compared to the deflection in the Schwarzschild limit. We observe that in the presence of uniform and non-uniform plasma distributions, there is a significant change in the deflection angle. Particularly, the presence of plasma increases the deflection of the light even if it has a uniform distribution. From Fig.~\ref{defang} one can also see that the inhomogeneity of the plasma adds some extra contribution to the change of the deflection angle. Fig.~\ref{defang2} shows the dependence of deflection angle on the quadrupole correction parameter for the fixed value of impact parameter $b$. From this dependence, one can see that the change of the deflection angle with respect to the parameter $ \epsilon $ is small in comparison to the changes due to the plasma. However, if we fix the plasma parameters, we can observe that for the increase of the module of negative values of $\epsilon$ the deflection angle increases.

\section{Lens equation and magnification\label{sect:magnification}}

In this section, we shall calculate the magnification of the source brightness using the formula for the deflection angle. Firstly, we shall consider the vacuum case, i.e. without the influence of plasma. Later we shall observe the plasma effects in the magnification.

We simply consider the quadrupole correction parameter $ \epsilon $ in the deflection angle equation. We consider the well known lens equation~\cite{Synge60,Schneider92, Morozova13}
\begin{equation}\label{lenseq}
\theta D_{s}=\beta D_{s}+\alpha D_{ls}\ .
\end{equation}
where $ \beta $ is the angle of the source from the observer-lens axis, $ \theta $ is the angle of the apparent image of the source due to lensing with the deflection angle $ \alpha $, $ D_{s} $ and $ D_{ls} $ are the distances from the observer to the source and from the lens to the source, respectively \cite{Tsupko12}. 

The impact parameter $ b $ and the angle $ \theta $ are related by the relation $ b=D_{l}\theta $. Here $ D_{l} $ is the distance of the lens from the observer. Introducing the quantity $ F(\theta)=|\alpha_{b}|b=|\alpha_{b}(\theta)|D_{l}\theta $, we write Eq.~(\ref{lenseq}) as
\begin{equation}\label{lenseq2}
\beta=\theta-\frac{D_{ls}}{D_{s}}\frac{F(\theta)}{D_{l}}\frac{1}{\theta}\ .
\end{equation}
Eq.~(\ref{lenseq2}) gives us the position of the image of the source due to lensing. For $ \beta=0 $, we get the Einstein angle $ \theta_{0} $ and it corresponds to the case when the observer, the lens and the object are in the same straight line. The Einstein ring is given by the equation $ R_{0}=D_{l}\theta_{0} $. Usually, the Einstein angle is too small to be resolved by modern telescopes. However, the lensing effects by some astrophysical objects like a star or a stellar-mass black hole can be detectable because of the change in apparent brightness of the source (magnification) by non-vanishing quadupole moments and presence of plasma in the vicinity of the object. The magnification of the image brightness can be calculated from the formula
\begin{equation}\label{magnification}
\mu_{\Sigma}=\frac{I_{tot}}{I_{*}}=\sum\limits_{k}\left| \left(\frac{\theta_{k}}{\beta}\right)\left(\frac{d\theta_{k}}{d\beta}\right) \right| \ \ \ \ k=1,2,...,s
\end{equation}
where $ s $ is the number of images, $ I_{tot} $ and $ I_{*} $ represent the total brightness of the images and the unlensed brightness of the source, respectively, and $ k $ refers to the number of images. 

\begin{figure*}[t]
	\begin{center}
		\includegraphics[width=0.45\textwidth]{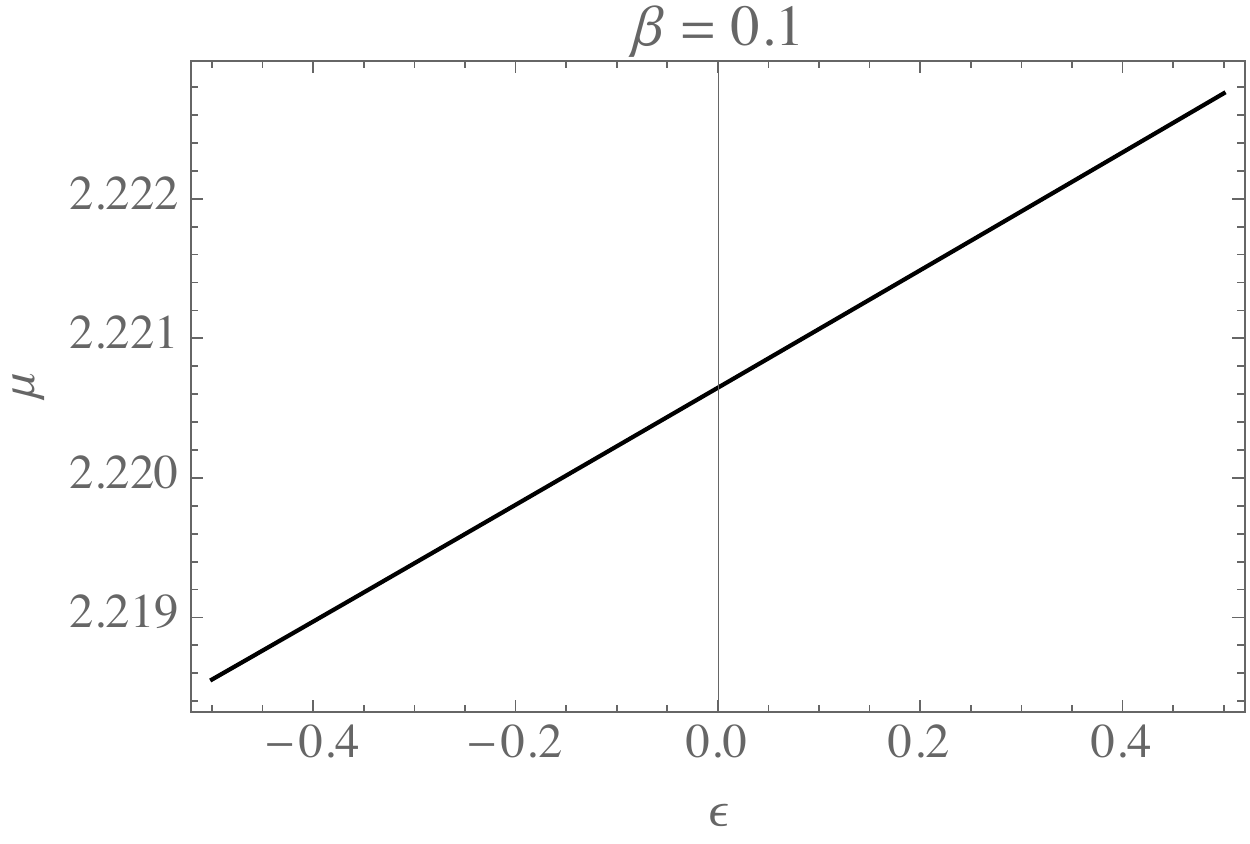}
		\includegraphics[width=0.45\textwidth]{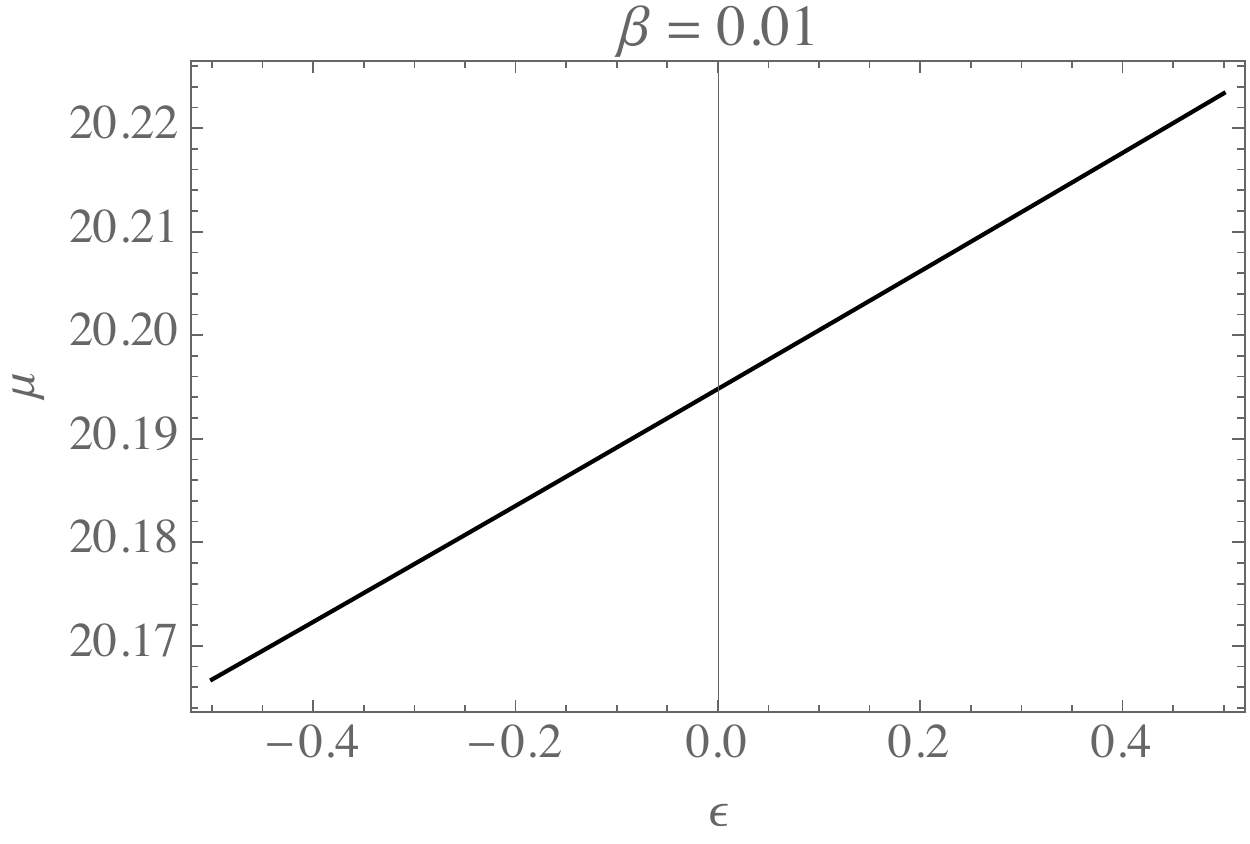}
	\end{center}
	\vspace{-0.5cm}
	\caption{The solid lines correspond to the total magnification under the effects of quarupole correction parameter $\epsilon$  \label{magwoplas}}
\end{figure*}

\subsection{Magnification without plasma}

Let us now consider the magnification of a non-rotating compact object without the presence of plasma. Ignoring the contribution from plasma, we write the deflection angle as
\begin{equation}
\hat{\alpha}_{b}=-\frac{4M}{b}+\frac{3\pi M^2}{b^{2}}+\frac{16\epsilon M^3}{3b^3}\ .
\end{equation}

Using Eq.~(\ref{numint2}), we write Eq.~(\ref{magnification}) as
\begin{equation}\label{eqa}
\beta=\theta-\frac{1}{\theta}\left[4M\left(1-\frac{3\pi M}{4D_{l}\theta}-\frac{4\epsilon M^2}{3 D_{l}^{2}\theta^{2}}\right) \right] \frac{D_{ls}}{D_{l}D_{s}}\ . 
\end{equation} 
Now using the notations 
\begin{equation}
\theta_{E}^{2}=4M\frac{D_{ls}}{D_{l}D_{s}}, \ \ \ \theta_{F}=\frac{3\pi M}{4D_{l}}\theta_{E}^{2}\ , \ \ \ \theta_{G}=\frac{4\epsilon M^{2}}{3D_{l}^{2}}\theta_{E}^{2}
\end{equation}
we can write (\ref{eqa}) as 
\begin{equation}\label{angleeq}
\theta^{4}-\beta\theta^{3}-\theta_{E}^{2}\theta^{2}+\theta_{F}\theta+\theta_{G}=0\ .
\end{equation}
The solutions of the lens equation~(\ref{angleeq})  give the positions $ \theta $ of  the  appeared  images  of  the  object  due  to  the presence of the lens. $ \theta_{E} $ in Eq.~(\ref{angleeq}) is called the Einstein angle for the Schwarzschild case and it sets the characteristic scale of lensing in the system. 

Now we try to solve Eq.~(\ref{angleeq}) to find the position of the images and subsequently we shall find the magnification by Eq.~(\ref{magnification}). Eq.~(\ref{angleeq}) has four distinct roots and these roots correspond to four different images. Using the solution for the lens equation, we write the magnification as,
\begin{equation}
\mu_{\Sigma}=\sum\limits_{k}\Big| \left(\frac{\theta_{k}}{\beta}\right) \left(\frac{d\theta_{k}}{d\beta}\right)  \Big|
\end{equation}
Now, the total magnification of the image source due to weak lensing in the non-rotating case can be represented as
\begin{equation}
\mu_{tot}=\mu_{+}^{(1)}+\mu_{-}^{(1)}+\mu_{+}^{(2)}+\mu_{-}^{(2)},
\end{equation}
where $ \mu^{(1)}_{\pm} $s' correspond to the first order magnification and in the limiting case, when $ \epsilon=0 $, they take the form of the Schwarzschild black hole case. $ \mu^{(2)}_{\pm} $ corresponds to the second order magnification and tends to vanish when $ \epsilon=0 $. We plot the total magnifications in Fig.~\ref{magwoplas} for $ \beta = 0.01 $ and $ \beta = 0.1 $. We observe that the change in the magnification is small, varying $ \epsilon $ for both $ \beta = 0.1 $ and $ \beta = 0.01 $. So it is extremely difficult to differentiate the effect of $ \epsilon $ on magnification observationally. But one can observe the number of images produced by the lensing system, which can suggest a possible modification due to quadrupolar correction. Gravitational lensing system with four images have been reported in \cite{Er14,Bradac02}. Other observations with highly magnified image has been reported in \cite{Ebeling18}.

\subsection{Magnification with Uniform Plasma Distribution}

In this subsection, we shall consider the image magnification by a non-rotating compact object in the presence of uniform plasma. Assuming $ \omega_{p}/\omega=0.5 $ in Eq.~(\ref{numint2}), the deflection angle becomes
\begin{equation}
\hat{\alpha}_{b}=-\frac{51M}{6b}+\frac{13\pi M^{2}}{4b^{2}}+\frac{35\epsilon M{3}}{6b^{3}}\ .
\end{equation}

With this deflection angle, the lens equation becomes
\begin{equation}\label{polynomp}
\theta^{4}-\beta\theta^{3}-\theta_{H}^{2}\theta^{2}+\theta_{I}\theta+\theta_{J}=0\ ,
\end{equation}
where
\begin{equation}
\theta_{H}^{2}=\frac{51M}{6}\xi, \ \ \ \ \theta_{I}=\frac{13\pi M^{2}}{4D_{l}}\xi, \ \ \ \ \theta_{G}=\frac{35\epsilon M^{3}}{6D_{l}^{2}}\xi.
\end{equation}
and
\begin{equation}
\xi = \frac{D_{ls}}{D_{l}D_{s}}
\end{equation}
Even in this case, there will be four images and the expression for magnification can be written as 
\begin{equation}
\mu_{\Sigma}=\sum\limits_{k}\Big| \left(\frac{\theta_{k}}{\beta}\right) \left(\frac{d\theta_{k}}{d\beta}\right)  \Big|
\end{equation}
%

\begin{figure*}[t]
	\begin{center}
		\includegraphics[width=0.45\textwidth]{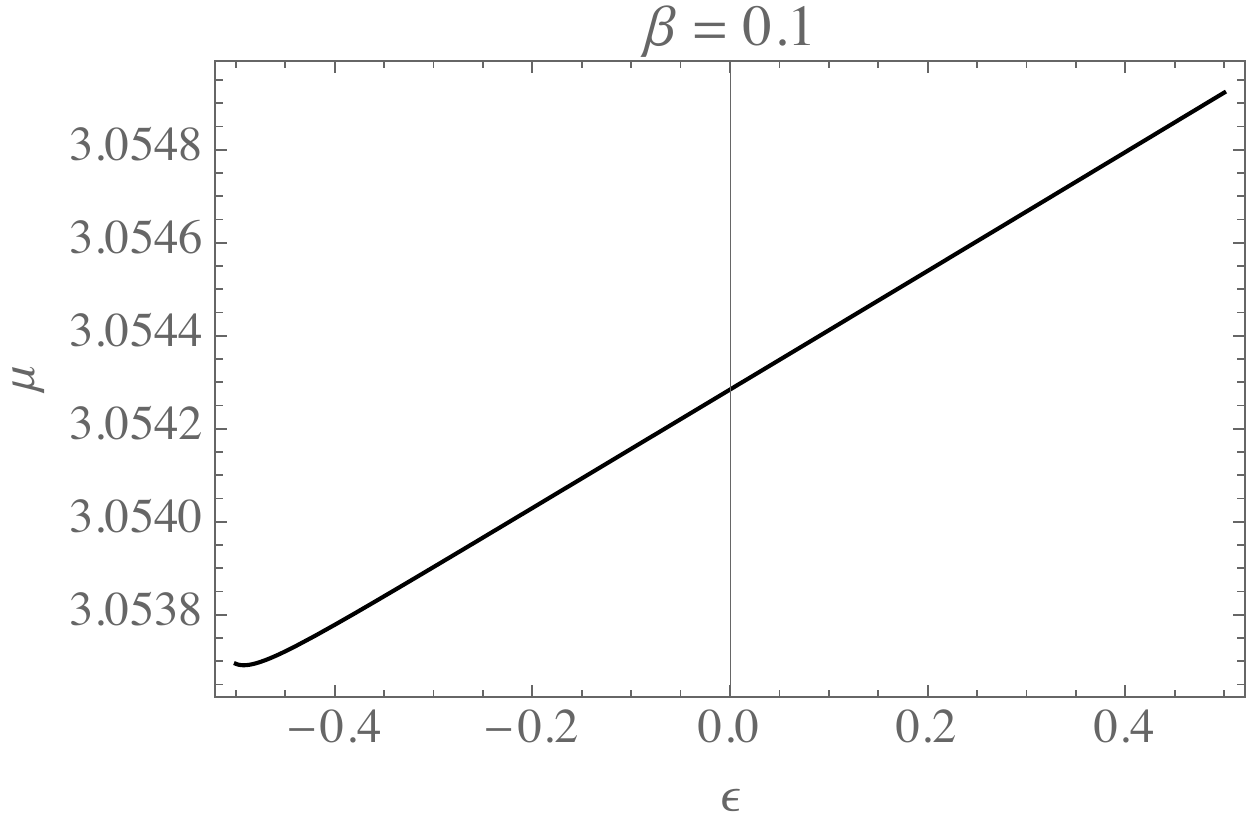}
		\includegraphics[width=0.45\textwidth]{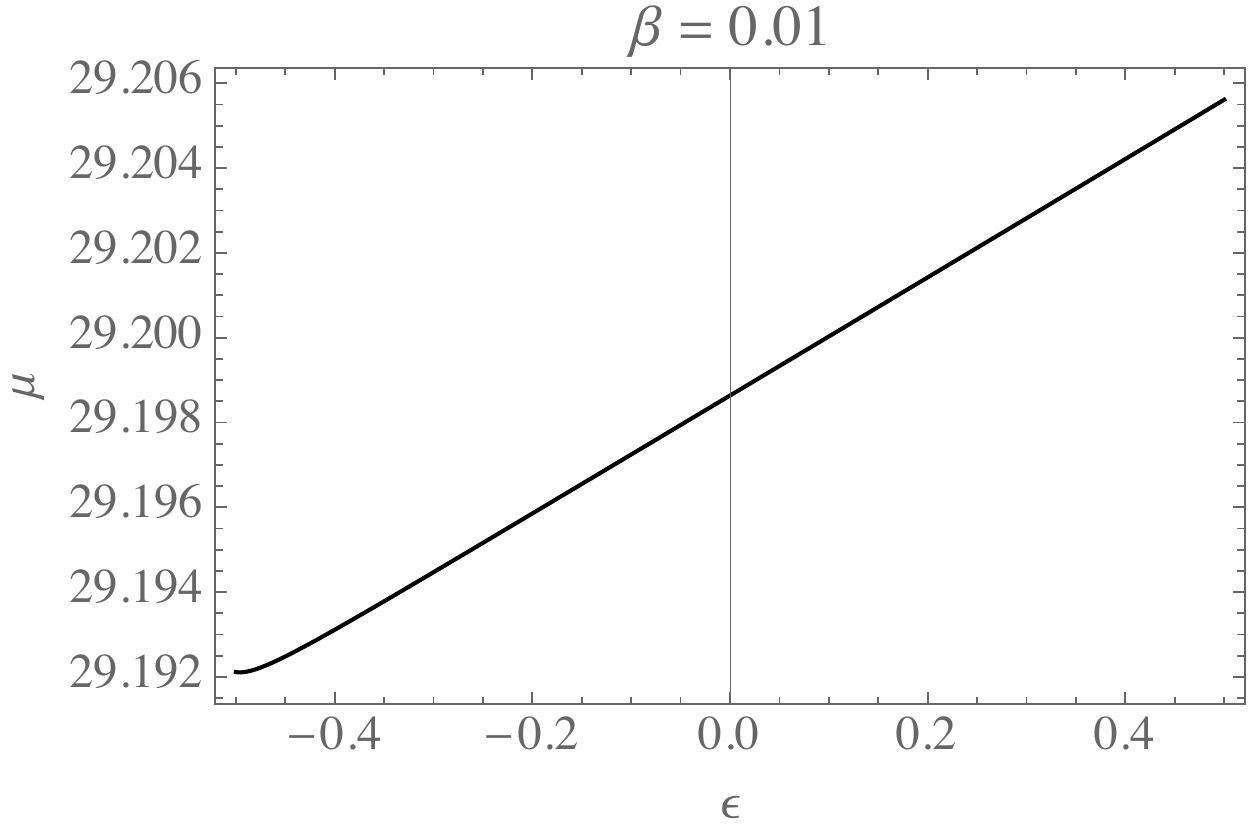}
	\end{center}
	\vspace{-0.5cm}
	\caption{The solid lines correspond to the total magnification under the effects of quarupole correction parameter $\epsilon$. \label{magwuplas}}
\end{figure*}

We solve the polynomial equation (\ref{polynomp}) for $ \theta_{k} $ and express the magnification as
\begin{equation}
\mu_{tot}=\mu_{+}^{(1)}+\mu_{-}^{(1)}+\mu^{(2)}_{+}+\mu_{-}^{(2)}
\end{equation}
Here $ \mu^{(1)}_{\pm} $s' correspond to the first order magnifications and in the limiting case, when $ \epsilon=0 $, they take the form of the Schwarzschild black hole case. The term $ \mu^{(2)}_{\pm} $ corresponds to the second order magnification with the effects of plasma and quadruple correction parameter. In Fig.~\ref{magwuplas}, we show the total magnification as a function of the quadruple correction parameter for $ \beta = 0.1 $ (left panel) and $ \beta = 0.01 $ (right panel). Similarly, in this case, we can see a change in the magnification by varying $ \epsilon $.   

\subsection{Magnification with Non-Uniform Plasma Distribution}

For Magnification of image by a non-rotating compact object in non-uniform plasma, we consider a plasma distribution of the form in Eq.~(\ref{plasmadist}). But for comparison we consider a power-law distribution of the form
\begin{equation}\label{density}
N(r)=\frac{N_{0}}{r^{h}}\ ,
\end{equation}
where, $ h>0 $. For numerical comparison we take $ h=3 $. 

Assuming $ k=k_{e}N_{0}=1 $ as an arbitrary constant, deflection angles takes the form
\begin{equation}\label{eqnonun}
\begin{aligned}
&\hat{\alpha}_{b}=-\frac{8M}{b}+\frac{3\pi M^{2}}{b^{2}}+\frac{2}{\omega^{2}b^{3}}+\frac{16\pi M^{3}}{3b^{3}} \\ 
& \ \ \ \ \ \ \ +\frac{32 M^{2}}{15\omega^{2}b^{5}}- \frac{3\pi M}{8\omega^{2} b^{4}}+ \frac{45\pi\epsilon M^{3}}{96\omega^{2}b^{6}}+ \frac{15\pi}{32\omega^{4}b^{6}}\ .
\end{aligned}
\end{equation}
Using the Eq.~(\ref{eqnonun}) the lens equation can be rewritten as 
\begin{equation}
\begin{aligned}
&\theta^{7}-\beta\theta^{6}-A\theta^{5}+B\theta^{4} \\ &+C\theta^{3}-D\theta^{2}+F\theta+G=0\ ,
\end{aligned}
\end{equation}
were A, B, C, D, F, and G, depends upon $ \epsilon $, $ M $ and the distances involved in the lensing system. The existence of real roots, for the particular values of the parameter $\epsilon$, non-uniform plasma distribution can produce seven different images of the lensed object, and the corresponding magnification of the source image brightness.


\section{Conclusion\label{sec:conclusion}}

In this paper, we have studied weak gravitational lensing around a compact object with arbitrary quadrupole moment in the presence of both uniform and nonuniform plasma. The results obtained here can be listed as follows:

\begin{enumerate}[label=\roman*)]

\item We presented the description of light propagation in homogeneous and non-homogeneous plasma media around compact object with arbitrary quadrupole moment in the weak field approximation. 

\item We studied the dependence of the deflection angle in weak gravitational lensing on the quadrupole correction parameter $ \epsilon $ encoded in the quasi-Kerr metric. We found that the presence of plasma can significantly modify the deflection angle but the effect of $ \epsilon $ is small. We found the linear term contributing to the deflection angle due to the new parameter $\epsilon$. 

\item We also studied the observable quantity magnification $ \mu $. We found that the effect of $ \epsilon $ on the magnification is not significant, but that of plasma is. However, the presence of the new parameter causes the appearance of additional images of the source and contributes to the magnification of brightness. 

\item In all the cases we considered non-homogeneous plasma with power-law distribution. This has a significant impact on the deflection angle and the lens equation becomes a seventh order polynomial. Such lensing systems can produce up to seven images of the source.

\end{enumerate}

As the next step of our study, we plan to consider the strong lensing effect around a quasi-Kerr black hole surrounded by plasma. We also plan to get the forms of the shadow cast this object and study the effects of the plasma on the change of its shape.


\begin{acknowledgments}

We thank Dimitry Ayzenberg for useful discussions and comments. This work was supported by the National Natural Science Foundation of China (Grant No.~U1531117) and Fudan University (Grant No.~IDH1512060). H.C. also acknowledges the support from the China Scholarship Council (CSC), Grant No.~2017GXZ019020. A.B.A. also acknowledges the support from the Shanghai Government Scholarship (SGS). A.A.A. acknowledges the support by Grant No. VA-FA-F-2-008 of the Uzbekistan Agency for Science and Technology. C.B. also acknowledges the support from the Alexander von Humboldt Foundation.
\end{acknowledgments}

\appendix 

\section{Quasi-Kerr spacetime}\label{s-review}

A rotating quasi-Kerr compact object with arbitrary quadrupole moment was proposed by Glampedakis and Babak~\cite{Glampedakis06b} with the help of the Hartle-Thorne spacetime. This metric has three independent parameters, i.e. the mass $ M $, the spin parameter $ a $, and the quadrupole correction parameter $ \epsilon $. The parameter $ \epsilon $ defines the deviation of the quadruple moment from that of a Kerr black hole. For the quasi-Kerr metric, the quadruple moment is $ q_{\rm Kerr}-\epsilon M^3 $, where $ q_{\rm Kerr}=-J^{2}/M $. The quadruple moment can be written as
\begin{equation} \label{qrplmom}
Q=-M(a^{2}+\epsilon M^{2})
\end{equation}
The Kerr metric in Boyer-Lindquist coordinate is written as

\be \label{kerr}
ds^{2}_{K}&=&-\Big(1-\frac{2Mr}{\Sigma}\Big)dt^{2}-\Big(\frac{4Mar\sin^{2}\theta}{\Sigma}\Big)dtd\phi +\frac{\Sigma}{\Delta}dr^{2}  \nonumber \\ &&+ \Sigma d\theta^{2}+\Big(r^{2}+a^{2}+\frac{2Ma^{2}r\sin^{2}\theta}{\Sigma}\Big)d\phi^{2}\ ,
\ee
where
\be
\Sigma &\equiv& r^{2}+a^{2}\cos^{2}\theta \\ \Delta &\equiv& r^{2}-2Mr+a^{2}\ . 
\ee
The quadrupolar correction is introduced by choosing a quadrupole moment of the form (\ref{qrplmom}) in the Hartle-Thorne metric. Then, the quasi-Kerr metric $ g^{qK}_{ab} $ in Boyer-Lindquist coordinate is given by~\cite{Glampedakis06b},
\be \label{qkmetric}
g^{qK}_{ab}=g^{K}_{ab}+\epsilon h_{ab}+{\cal{O}}(\delta M_{l\geq 4} ,\delta S_{l\geq3}) \, ,
\ee
where $ M_{l} $ and $ S_{l} $ are, respectively, the mass and the current multipole moments of order $ l $, where $l \geq 0$ is the angular integer eigenvalue. $ h_{ab} $s' are given by
\be
h^{tt}&=&(1-2M/r)^{-1}\left[(1-3\cos^{2}\theta)F_{1}(r)\right], \\
h^{rr}&=&(1-2M/r)\left[(1-3\cos^{2}\theta)F_{1}(r)\right],  \\
h^{\theta\theta}&=&-\frac{1}{r^{2}}\left[(1-3\cos^{2}\theta)F_{2}(r)\right],  \\
h^{\phi\phi}&=&-\frac{1}{r^{2}\sin^{2}\theta}\left[(1-3\cos^{2}\theta)F_{2}(r)\right], \\
h^{t\phi}&=&0\ ,
\ee
where the functions $ F_{1}(r) $ and $ F_{2}(r) $ have the following form  
\be 
F_{1}&=&-\frac{5(r-M)(2M^{2}+6Mr-3r^{2})}{8Mr(r-2M)}\nonumber \\
&& -\frac{15r(r-2M)}{16M^{2}}\ln\left(\frac{r}{r-2M}\right)  \ ,\\
F_{2}&=&\frac{5(2M^{2}-3Mr-3r^{2})}{8Mr} \nonumber \\
&&+\frac{15(r^{2}-2M^{2})}{16M^{2}}\ln\left(\frac{r}{r-2M}\right).
\ee

Now the line element has the form
\begin{eqnarray} \label{qkmetric1}
ds^{2} &=&g_{00}dt^2+g_{11}dr^{2}+g_{22}d\theta^{2}+g_{33}d\phi^{2}
\nonumber\\ && +2g_{03}dtd\phi
\end{eqnarray}
where
\begin{eqnarray}
g_{00}&=&-\left(1-\frac{2Mr}{\Sigma}\right)+\epsilon(1-3\cos^{2}\theta) \nonumber\\
&&\times \left(\frac{F_{1}\left(1-{2Mr}/{\Sigma}\right)^{2}}{1-{2M}/{r}}+\frac{4a^{2}F_{2}M^{2}\sin^{2}\theta}{\Sigma^{2}}\right), \\
g_{11}&=&\frac{\Sigma}{\Delta}+\epsilon \frac{F_{1}(1-{2M}/{r})\Sigma^{2}(1-3\cos^{2}\theta)}{\Delta^{2}}, \\
g_{22}&=&\Sigma - \frac{F_{2}\Sigma^{2}(1-3\cos^{2}\theta)}{r^{2}},  \\
g_{33}&=&\left(r^{2}+a^{2}+\frac{2a^{2}Mr\sin^{2}\theta}{\Sigma}\right)\sin^{2}\theta \nonumber \\
&&+\epsilon(1-3\cos^{2}\theta)\sin^{2}\theta \Big[\frac{4a^{2}F_{1}M^{2}r^{2}}{\left(1-{2M}/{r}\right)\Sigma^{2}} \nonumber\\ &&-\frac{F_{2}}{r^{2}}\left(r^{2}+a^{2}+\frac{2a^{2}Mr\sin^{2}\theta}{\Sigma}\right)^{2}\Big], \\
g_{03}&=&-\frac{2aMr}{\Sigma}\sin^{2}\theta+\epsilon (1-3\cos^{2}\theta)\sin^{2}\theta \nonumber \\ 
&&\times \Bigg[\frac{2aF_{1}Mr\left(1-{2M}/{r}\right)}{\left(1-{2M}/{r}\right)\Sigma} \nonumber \\ 
&&+\frac{2aF_{2}M}{r\Sigma}\left(r^{2}+a^{2}+\frac{2a^{2}Mr\sin^{2}\theta}{\Sigma}\right)\Bigg],
\end{eqnarray}				
with $ \Sigma=r^{2}+a^{2}\cos^{2}\theta $ and $ \Delta=r^{2}-2Mr+a^{2} $. Note that the quasi-Kerr metric reduces to the Kerr one for $ \epsilon=0$.


\bibliography{gravreferences}

\end{document}